\documentclass[12pt, superscriptaddress, a4paper, reprint]{revtex4-2}
\usepackage[utf8]{inputenc}
\usepackage{amsmath}
\usepackage{graphicx, color}
\usepackage[colorlinks]{hyperref}
\usepackage{bibunits}

\begin{document}
\title{Few-emitter lasing in single ultra-small nanocavities}

\author{Oluwafemi S. Ojambati}
\affiliation{NanoPhotonics Centre, Cavendish Laboratory, Department of Physics, JJ Thompson Avenue, University of Cambridge, Cambridge, CB3 0HE, United Kingdom}
\affiliation{Present address: Nanobiophysics, MESA+ Institute for Nanotechnology, Faculty of Science and Technology, University of Twente, Enschede, The Netherlands}
\author{Krist{\'i}n B. Arnard{\'o}ttir}
\author{Brendon W. Lovett}
\author{Jonathan Keeling}
\affiliation{SUPA, School of Physics and Astronomy, University of St Andrews, St Andrews KY16 9SS, United Kingdom}
\author{Jeremy J. Baumberg}
\affiliation{NanoPhotonics Centre, Cavendish Laboratory, Department of Physics, JJ Thompson Avenue, University of Cambridge, Cambridge, CB3 0HE, United Kingdom}

\begin{abstract}
Lasers are ubiquitous for information storage, processing, communications, sensing, biological research, and medical applications~\cite{eichler2018lasers}.
To decrease their energy and materials usage, a key quest is to miniaturize lasers down to nanocavities~\cite{Ma_NatNano_19}. 
Obtaining the smallest mode volumes demands plasmonic nanocavities, but for these, gain comes from only single or few  emitters. Until now, lasing in such devices was unobtainable due to low gain and high cavity losses~\cite{Hill_NatPhoton_14}. 
Here, we demonstrate a plasmonic nanolaser approaching the single-molecule emitter regime. 
The lasing transition significantly broadens, and depends on the number of molecules and their individual locations. 
We show this can be understood by developing a theoretical approach~\cite{Arnardottir_PRL_20} extending previous weak-coupling theories~\cite{Rice_PRA_94}.
Our work paves the way for developing nanolaser applications~\cite{Ma_NatNano_19, Galanzha_NatComm_17, Kravets_Chev_18} as well as fundamental studies at the limit of few emitters~\cite{Rice_PRA_94, Jones1999Photon,Chow_APR_18}.
\end{abstract}

\maketitle

Lasing occurs when stimulated emission into a cavity mode exceeds loss, leading to amplification.  Typically, this causes a sharp change of slope in the emission \textit{vs} input power.  Such a sharp transition is analogous to a thermodynamic phase transition~\cite{graham1970laserlight}. 
As with phase transitions, there exists a `system-size' parameter $\beta$, which determines how sharp the transition is~\cite{Rice_PRA_94}.
However, questions remain about the sharpness of this transition
for lasers with few emitters, small cavities, and stronger light-matter coupling.   Such questions are particularly important for nanocavities which confine light within sub-wavelength volumes~\cite{Koenderink_Sci_15}, such as metasurfaces or plasmonic nanocavities that exploit collective electron oscillations in metallic nanostructures to achieve extreme confinement ($V<100\, \rm{nm}^3$)~\cite{Meinzer_NatPhoton_14, Baumberg_NatMat_19}. These enable coupling single emitters to light~\cite{Nomura_OptExp_09, Chikkaraddy_Nature_16, Ojambati_NatCom_19}.  Lasing in such plasmonic nanocavities presents new opportunities for miniaturisation and integration, but also raises new questions about the nature and conditions required for lasing in a regime which combines a small number of strongly-coupled emitters with lower quality resonators, $Q\sim 10$.  Our aim here is to understand this regime.

\begin{figure}[h]
    \centering
    \includegraphics[scale = 0.37]{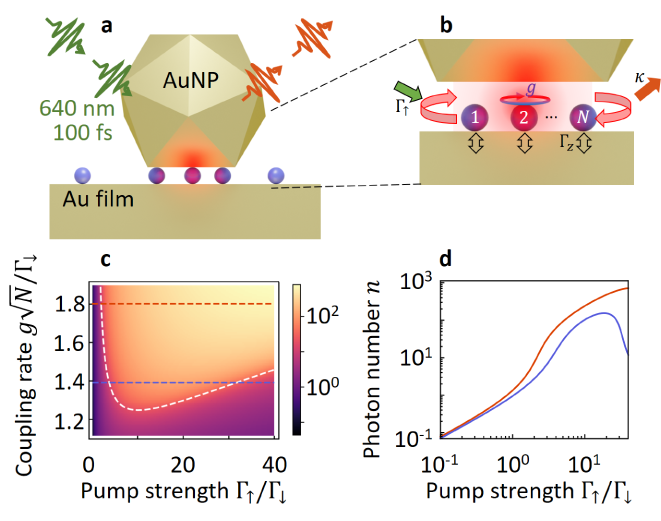}
    \caption{\label{fig:fig1}\textbf{Plasmonic nanocavity with emitters.} \textbf{a}, Nanoparticle-on-mirror (NPoM) cavity, formed by Au nanoparticle trapping few emitters in 0.9~nm gap above Au film. \textbf{b}, Nanogap region with $N$ emitters of dephasing $\Gamma_z$ excited by pump strength $\Gamma_\uparrow$, emitting into cavity with loss rate $\kappa$. \textbf{c}, Re-entrant lasing threshold, shown as colormap of photon number (yellow=high, purple=low) $vs$ normalized coupling and pump strength, using $\kappa=0.1\Gamma_\downarrow$, $N=10$, and $\Gamma_z=10\Gamma_\downarrow$. Dashed white line shows threshold defined by Eq.~\eqref{Eq:Gammac}. \textbf{d}, Cross sections, as indicated by horizontal dashed lines in (c). At large coupling the traditional input--output curve is seen (red line) while at smaller coupling lasing is suppressed at higher pump power (blue line).
    }
\end{figure}

The sharpness of the lasing transition reflects how lasing enhances the conversion efficiency of input  power into output light.  Above threshold the efficiency is high, as stimulated emission directs almost all radiation  into the cavity mode.  A sharp transition requires low efficiency below threshold.  This is captured by the parameter $\beta$,  the ratio of input-output slopes below and above threshold. Small $\beta$ indicates a sharp transition.  
When the light-matter coupling $g$ is weak,  $\beta=g^2/(g^2 + \Gamma_\downarrow\Gamma_T)$, where $\Gamma_T$ is the total emitter linewidth, and $\Gamma_\downarrow$ the decay rate into non-cavity modes~\cite{Rice_PRA_94}.
Below threshold, $\beta$ is the fraction of incoherent emission into the cavity and in this weak-coupling case, it fully determines the shape of the input-output curve.
It is notable that $\beta$ does not depend on the number of emitters, whereas  the parameter determining the sharpness of thermodynamic phase transitions is typically the system size. As we discuss below, allowing for strong light-matter coupling changes this behaviour considerably.

\begin{figure*}[ht]
    \centering
    \includegraphics[scale = 0.44]{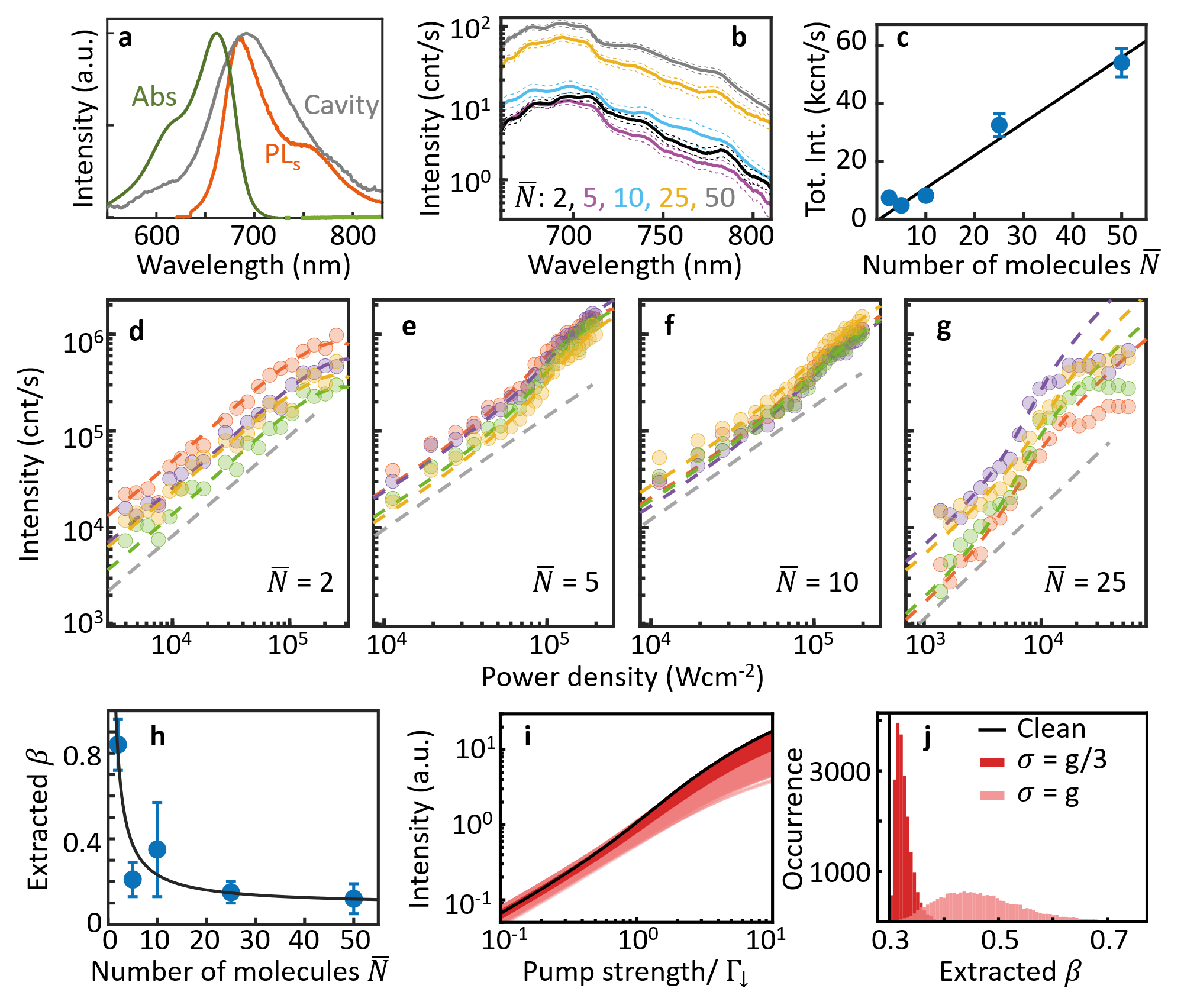}
    \caption{\label{fig:fig2} \textbf{Emission dependence on excitation power and number of emitters.} \textbf{a}, Absorption (green) and emission (orange) spectra of methylene blue in solution, and typical NPoM darkfield scattering spectrum (grey). \textbf{b}, Dashed lines: Example emission spectra for different expected number of emitters $\bar{N}$ in NPoM. Solid curves are averaged over $>$ 50 NPoMs. \textbf{c}, Spectral integrated intensity $vs$ $\bar{N}$ (blue circles) and linear fit (solid curve). Error bar is the standard error. \textbf{d-g}, Spectral integrated intensity $vs$ power density for different $\bar{N}$, colours correspond to different NPoMs. Dashed coloured lines are theoretical fits (see Methods) and the dashed grey lines are guide-to-the eye linear trends, which clearly deviate from the $\bar{N}>2$ data. \textbf{h}, Extracted slope ratio $\beta$ $vs$ $\bar{N}$ with predicted theoretical trend (solid curve), bars give range. \textbf{i,j}, Theoretical input-output curves, and extracted distribution of $\beta$, showing effect of disorder in couplings $g_i$.  Black line shows the homogeneous limit, $g_i=g=1.5 \Gamma_\downarrow$ for $N=10$ emitters, with $\Gamma_z=\Gamma_\downarrow, \kappa=1.74\Gamma_\downarrow$.  The red and pink curves (and corresponding distributions of $\beta$) arise by sampling $g_i$ from a Gaussian distribution with root mean square $g=1.5\Gamma_\downarrow$, and standard deviation $g/3, g$ respectively.The thick black line corresponds to the homogeneous case, where no disorder is present.
    }
\end{figure*}

Some extreme limits of lasing have previously been explored.  The single-emitter limit~\cite{Jones1999Photon} has been studied with atoms~\cite{mckeever2003experimental}, superconducting circuits~\cite{astafiev2007single}, and quantum dots~\cite{Laussy2008,Nomura_OptExp_09}.  
In this limit, the transition broadens as $\beta$ becomes large. This is because lasing with $N$ emitters requires $NC>1$, where the cooperativity  $C=4g^2/(\kappa\Gamma_T)$ depends also on the cavity loss rate  $\kappa$.  When $N=1$, one needs $C>1$, and so a sharp lasing threshold ($\beta \ll 1$) is only possible if $\kappa\ll\Gamma_\downarrow$, which is not typically the case.  

In this paper we explore lasing  of a few organic molecules coupled to a plasmonic nanocavity.  Despite low $Q$, emitters coupled to plasmonic modes can achieve lasing~\cite{Ma_NatNano_19}. However the smallest lasers must lase with only a few emitters, a goal so far thwarted, but attainable by using ultrasmall volume plasmonic nanocavities. These can be realised using bottom-up self-assembly, which we acheive via the nanoparticle-on-mirror (NPoM) geometry (Fig.~\ref{fig:fig1}a) of a Au nanoparticle on a thin dielectric spacer above an Au film~\cite{Baumberg_NatMat_19}.
We trap light inside 1~nm-high $\sim$20~nm-wide gaps, which enhance incident optical fields by $>$250 while retaining $\sim 50\%$ radiative efficiency. 
Enhancing both the pumping and the light-matter coupling  now enables lasing, confirmed by enhanced coherence in the lasing state.
To describe such experiments we must extend previous theoretical treatments~\cite{Rice_PRA_94} to  consider the combination of stronger coupling, few emitters, bad cavities, pumping-induced noise, and inhomogeneity.   

We model the light-matter interaction in this NPoM nanocavity using a standard master equation  describing many two-level molecules coupled to light with strength $g_i$.  The molecules undergo incoherent pumping, loss, and pure dephasing at rates $\Gamma_\uparrow, \Gamma_\downarrow ,\Gamma_z$ (see Methods, Fig. 1b). We then make a second-order cumulant expansion~\cite{Arnardottir_PRL_20}, giving coupled equations for photon number $n$, molecule-photon coherence, inter-molecular coherence, and molecular inversion. This approach is ideal for understanding how system size $N$ controls the sharpness of the transition, since it captures finite size effects treating $N$ as a parameter. It also includes the effects of spontaneous emission, and recovers the semiclassical theory of lasing~\cite{haken1970semiclassical, Rice_PRA_94} in appropriate limits.  
In contrast to weakly-coupled models of incoherent emission and absorption~\cite{Rice_PRA_94} our approach captures the back-action of the coherent light  on the dynamics of the emitters.

Considering first the homogeneous case where $g_i=g$, we solve for steady-state $n$ with constant pump $\Gamma_\uparrow$ giving
\begin{equation}
        n = \frac{n_0}{2}\left[\left(\frac{\Gamma_\uparrow}{\Gamma_c}-1\right) +\sqrt{\left(\frac{\Gamma_\uparrow}{\Gamma_c}-1\right)^2 + 4\beta(\Gamma_\uparrow) \frac{\Gamma_\uparrow}{\Gamma_c}}\right]\label{Eq:nbeta}
\end{equation}
with $n_0$ defined in the methods, and
\begin{align}
    \beta(\Gamma_\uparrow) &=\frac{2\kappa\Gamma_T}{N(\kappa+\Gamma_T)}\frac{(NC)^2}{(NC-1)^2\Gamma_c},\label{Eq:beta}\\
    \Gamma_c &=\frac{NC+1}{NC-1}\Gamma_\downarrow + \frac{NC}{NC-1}\frac{\kappa\Gamma_T}{N(\kappa+\Gamma_T)},\label{Eq:Gammac}
\end{align} 
where the total linewidth is $\Gamma_T=\Gamma_\uparrow+\Gamma_\downarrow+4\Gamma_z$.  

While Eq.~\eqref{Eq:nbeta} appears identical to that found in~\cite{Rice_PRA_94}, a subtle and crucial difference exists: the terms $\Gamma_c$ and $\beta$  both implicitly depend on $\Gamma_\uparrow$, because the cooperativity $C$ depends on  $\Gamma_T$.
This $\Gamma_\uparrow$ dependence can often be ignored, when $\Gamma_T \simeq  4\Gamma_z$. However in the regime where the collective cooperativity is low (as here), it is important to keep this dependence. 
Physically, strong pumping introduces phase noise which ultimately kills lasing by suppressing the cooperativity.  
For typical lasers this effect only occurs at very high powers, however for plasmonic nanocavities with high losses, a switch-on and switch-off transition are seen, and these can merge where $NC \simeq 1$~\cite{bohnet2012steady}.
This is seen in Fig.~\ref{fig:fig1}(c,d), which show how the photon number depends on the normalised pump power and coupling strength.  

A second crucial difference from~\cite{Rice_PRA_94} is the form of $\beta$ in Eq.~\ref{Eq:beta}.  In particular, this now depends on $N$, with $\beta \sim 1/N$ at large $N$.  This encodes the anticipated effect that larger systems show smaller finite-size effects.

Experimentally, NPoMs with $\sim$0.9~nm gap are defined by a monolayer of cucurbit[7]uril molecules that can each encapsulate a single molecule of methylene blue~\cite{Chikkaraddy_Nature_16} (for sample preparation see Methods). We vary the expected number of emitters $\bar{N}$ in the nanocavity by changing the ratio of emitting methylene blue to non-emitting cucurbit[7]uril molecules. Because the location of the emitting molecules is different in each NPoM, their overlap with the nanocavity mode and hence coupling $g_i$ varies. Using NPoM nanoparticle diameters of 60~nm tunes the dominant NPoM cavity mode to the gain peak (Fig.~\ref{fig:fig2}a), optimising the interaction of the emitted light with the nanocavity.

\begin{figure*}[ht]
    \centering
    \includegraphics[scale = 0.4]{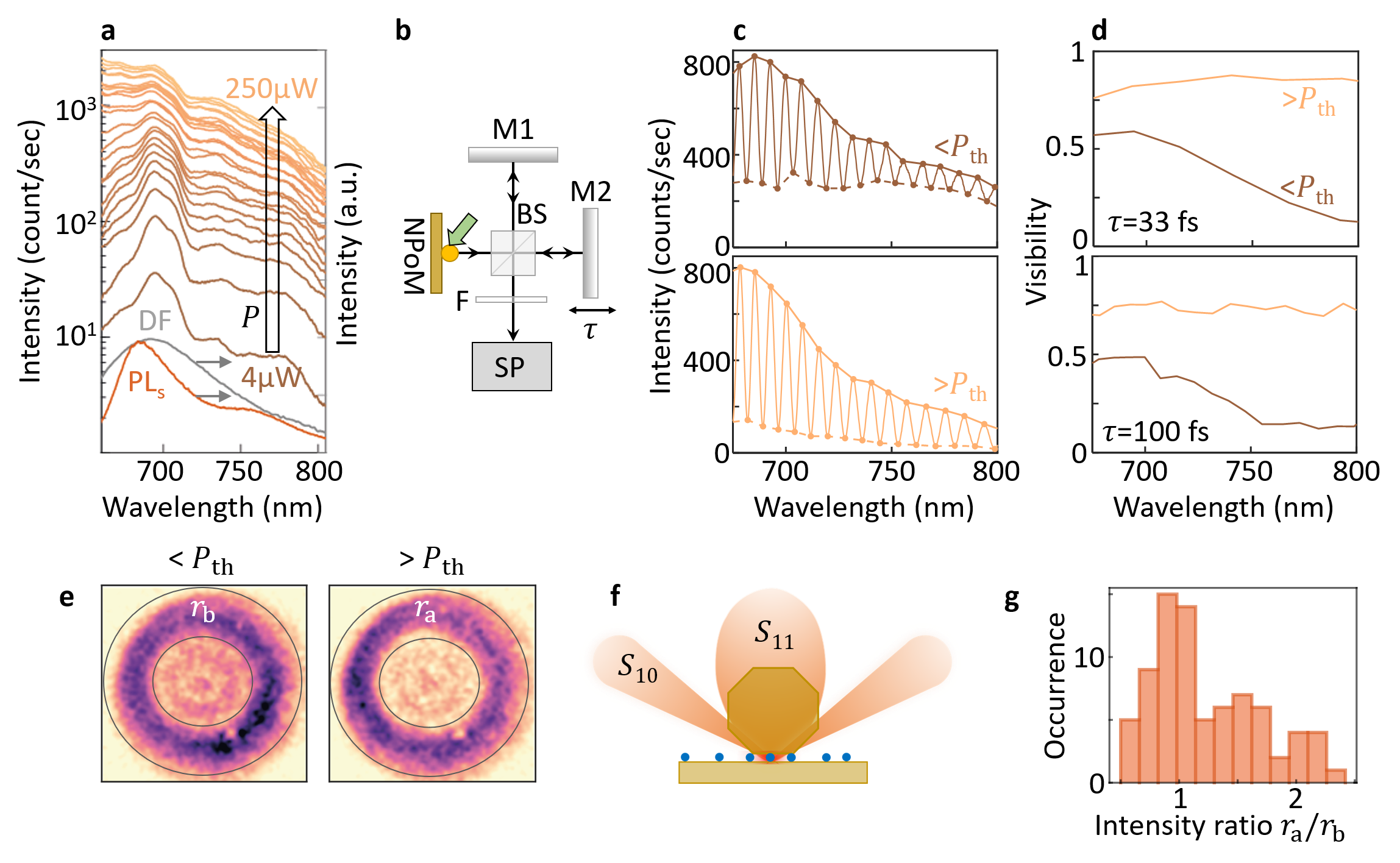}
    \caption{\label{fig:fig3} \textbf{Emission spectra and coherence of nanolaser}. \textbf{a}, Emission spectra for increasing excitation powers $P$, with dark-field (DF) and solution PL. \textbf{b}, Schematic of Michelson interferometer to measure spectral coherence. Emission from NPoM is split in two by 50:50 beamsplitter (BS), reflected from mirrors (M1,M2) with delay in one arm and recombined, filtered (F) and sent to spectrometer (SP). \textbf{c}, Spectral fringes with average powers below (upper panel) and above (lower panel) lasing threshold ($P_{\mathrm{th}}$).  \textbf{d}, Visibility \textit{vs} wavelength at different delays $\tau$. \textbf{e}, Far-field Fourier space images of emission below and above threshold, divided into annular ring and centre.  The integrated intensities in these regions, $I_{\text{ring}}$ and $I_{\text{centre}}$ allow us to define an intensity ratio $r=I_{\text{ring}}/I_{\text{centre}}$ both below threshold ($r_{\mathrm{b}}$) and above threshold ($r_{\mathrm{a}}$).
    Numerical aperture 0.9 limits collection angle $< 64^{\circ}$. \textbf{f}, NPoM radiative modes $S_{10}$ and $S_{11}$ that emit at high and low angles, respectively. \textbf{g}, Histogram of relative intensities in the high-angle ring $r_{\mathrm{a}}/r_{\mathrm{b}}$. }
    
\end{figure*}

To create enough gain, ultrafast pulses (100~fs, 640~nm) are used to irradiate individual NPoM cavities whose emission is recorded by a low-noise spectrometer (for experimental setup see Methods). At average powers of $P_{\rm{av}} = 4\,\mu$W (power density = $3.2\,\rm{kWcm^{-2}}$), the emission spectra for different numbers of molecules is similar and the total integrated intensity varies linearly with $\bar{N}$ (Fig.~\ref{fig:fig2}b,c).

Power dependent measurements on multiple NPoM constructs are performed with different $\bar{N}$ (Fig.~\ref{fig:fig2}d-g). For NPoM cavities with lower $\bar{N}$, smooth transitions in the total emitted intensity are seen, becoming systematically sharper for higher $\bar{N}$. In particular for $\bar{N}\sim2$, no observable transition is seen and the total emission is linear before saturating at the highest powers. This is consistent with thresholdless lasing, approaching the single-emitter regime. For NPoMs with large $\bar{N}$, well-pronounced thresholds are seen with super-linear emission, which then saturate to linear scaling as in conventional lasers. We fit Eq.~($\ref{Eq:nbeta}$) to the experimental data (for fitting procedure see Methods);  our theory matches well across the range of $\bar{N}$, indicating lasing in plasmonic nanocavities with a few emitters~\footnote{The deviation seen at large powers for $\bar{N}=25$ could result from intermolecular quenching.}. Furthermore, the extracted $\beta$-factor varies inversely with $\bar{N}$ (Fig.~\ref{fig:fig2}h), as expected from our theoretical model (Eq.~\ref{Eq:beta}).

The analytic form of Eq.~\eqref{Eq:nbeta} ignores the effects of random placement of molecules, which gives variable light-matter coupling.
This variation becomes important when the number of molecules is small, as it leads to significant variation between different NPoMs with the same number of molecules.  With random positions one finds that the shape of the input-output curve (and thus the parameter $\beta$) depends on the distribution of couplings, as shown in Fig.~\ref{fig:fig2}(i,j).


Because the threshold is less defined at small $N$, it becomes important to find further experimental evidence for lasing.  To this end, we examine the spectral evolution and coherence across the transition. As the average power $P_\mathrm{av}$ increases from 4~$\mu$W to 250~$\mu$W (Fig.~\ref{fig:fig3}a), the emission (which is 10~nm red-shifted from the solution PL due to the cavity environment) significantly broadens, with increasing weight at shorter (bluer) wavelengths. This behaviour appears consistently in most NPoMs at all $\bar{N}$, with emission even extending to the blue side of the pump laser wavelength (see SI). 
One explanation for this change in emission spectrum with power is a breakdown of Kasha's rule~\cite{kasha1950characterization}.  Kasha's rule states photoluminescence is only from the lowest energy vibrational state, as excited states relax before luminescence.
However, radiative decay is drastically sped up in NPoMs, reaching below $50\,$fs as the Purcell factor is $\sim 10^5$ due to the large $Q/V$. When emission becomes stimulated, this can become faster than the vibrational lifetime ($\sim 1\,$ps) \cite{Ojambati_NatCom_19}, allowing direct emission from higher vibronic states, breaking Kasha's rule~\cite{DelPino2018Tensor}.

To track the coherence of the observed emission we explore spectral interference using a Michelson interferometer (Fig.~\ref{fig:fig3}b). The delay time $\tau$ between the two arms of the interferometer is greater than the temporal coherence time ($\tau > \tau_\mathrm{c} \sim \hbar/\Gamma_T \sim 13\,$fs). Spectral fringes are observed both below and above threshold (Fig.~\ref{fig:fig3}) with a wavelength period $7~$nm inversely proportional to $\tau$. The extracted  visibility from the spectral envelope of the fringes, decreases with wavelength below $P_\mathrm{th}$ to 0.1 at 800~nm (shown for two different time delays $\tau = \{33,100\}$~fs, in Fig.~\ref{fig:fig3}c). By contrast above threshold, visibility increases to 0.8 becoming constant over the entire spectrum.


As discussed in Ref.~\cite{Yang2006Coherence}, by combining spectral filtering with coherence measurements, we show in Fig~\ref{fig:fig3}c,d how the spatial coherence of the light changes through the threshold.
To better understand this increasing visibility, the far-field emission is measured in Fourier space (for optical setup see Methods). The dominant bright plasmonic mode $S_{10}$ emits at high angles both below and above threshold (seen as bright, purple-coloured, ring in Fig.~\ref{fig:fig3}e), around $40-50^{\circ}$ (Fig. \ref{fig:fig3}f). Additional weaker emission at low angles---arising from the $S_{11}$ mode~\cite{Horton_PNAS_2020}---is proportionately larger below threshold. This mode mostly outcouples light from molecules  near the edges of the nanoparticle facet (Fig. \ref{fig:fig3}f), but only with poor quantum efficiency $<5\%$\cite{Horton_PNAS_2020}.
We define the emission ratio $r=I_{\text{ring}}/I_{\text{centre}}$ between the intensity from the annular ring $I_{\text{ring}}$ and from the centre $I_{\text{centre}}$. This  measures the relative emission of these modes.
By comparing the emission ratios above and below threshold $r_\mathrm{a,b}$, we see that $r_\mathrm{a}/r_\mathrm{b} \sim$1 for some NPoMs but $>$1 for many others. 
This indicates preferential occupation of one mode, as is typical in lasers, due to stimulated emission.


In summary, we show extreme confinement of optical fields exciting a few molecules in a nanometer gap that reveals lasing is possible with a few emitters, and that this can match only extended models.
This detailed influence of cavity $Q$ allows for future work exploring the photon statistics from such ultimately miniaturised lasers, and clarifying the fundamentals of the lasing nonlinear phase transition.


\bibliography{references}

\newpage
\twocolumngrid 
 
\section*{Methods}

\subsection{Theoretical modelling}

In this section we describe the theoretical approach we use to calculate the input-output curve for lasing with few strongly-coupled emitters.   We first introduce the model, then discuss the cumulant approach, and how the resulting equations can be solved with and without disorder in the matter-light coupling.

\subsubsection{Model}

We model the system as a plasmonic mode coupled to a collection of $N$ two-level systems: 
\begin{align}
    H=\omega a^\dagger a + \sum_i^N \left[ \frac{\epsilon}{2}\sigma_i^z + g_i(a^\dagger\sigma_i^- + a\sigma_i^+)\right],
\end{align} 
here, $a^{(\dagger)}$ is the annihilation (creation) operator for the plasmonic mode of energy $\omega$ while $\sigma^\alpha_i$ are the Pauli matrices representing the two-level system at site $i$.
We use a Lindblad master equation to account for the following incoherent processes with their corresponding rates: photon loss ($\kappa$), incoherent pump ($\Gamma_\uparrow$), non-radiative losses ($\Gamma_\downarrow$) and pure dephasing ($\Gamma_z$).  We thus have
\begin{multline}
    \partial_t \rho = -i[H,\rho] + \kappa \mathcal{D}[a] 
    \\+ \sum_i\left( \Gamma_\downarrow\mathcal{D}[\sigma_i^-] +
    \Gamma_\uparrow\mathcal{D}[\sigma_i^+] +
    \Gamma_z\mathcal{D}[\sigma_i^z]
    \right),
\end{multline}
with $\mathcal{D}[X]=X\rho X^\dagger + \frac{1}{2}\left(X^\dagger X\rho + \rho X^\dagger X\right)$.

The model we consider does not explicitly include the vibronic progression present in organic molecules~\cite{Arnardottir_PRL_20}.  The incoherent pumping rate $\Gamma_\uparrow$ we consider should however be understood as an effective rate, arising from the combination of coherent pumping exciting a higher vibrational state followed by vibrational relaxation (see Fig.~\ref{fig:fig1}).  This simplified model is chosen since, as discussed below, it allows for closed-form expressions for the lasing threhsold.  Exploring how more complex treatment of the molecular spectrum affects these results is an important question for future work.

\subsubsection{Cumulant equations}

Various approaches exist to model a system described by the above equations~\cite{Kirton2019Introduction}.  
In this work, as discussed in the main text, our focus is on understanding how system size $N$ enters into determining the sharpness of the transition.  For this, it is useful to choose an approach which captures finite size effects, treats $N$ as a parameter, and captures the semiclassical effects of spontaneous emission.  The ideal approach for this is to use second order cumulants~\cite{Kirton2017Suppressing,Arnardottir_PRL_20}, which provide a systematic expansion in $1/N$.


From the density matrix equation of motion given above, we can write down equations of motion for the second order moments of the system, using cummlant expansion to break higher order moments into a combinations of first and second order moments. The non-zero moments are:  $n=\langle a^\dagger a\rangle$, $P_i=\text{Im}\langle a\sigma_i^+\rangle, S_i = \langle \sigma_i^z\rangle$ and $D_{ij}=\langle\sigma_i^+\sigma_j^-\rangle,$ where we require the two operators in the last moment to act on different molecules. Assuming the resonant case $\omega=\epsilon$, the equations of motion end up taking the form:
\begin{align}
    \partial_t n &= -\kappa n - 2\sum_i g_i P_i\label{Eq:dn};\\
    \partial_t P_i &= -\frac{\kappa+\Gamma_T}{2}P_i -
                g_i\left(nS_i \frac{S_i+1}{2} + \sum_{j\neq i}D_{ij}\right)\label{Eq:dP};\\
    \partial_t S_i &= - (\Gamma_\uparrow+\Gamma_\downarrow)(S_i-S_0) + 4g_iP_i\label{Eq:dS};\\
    \partial_t D_{ij} &= -\Gamma_TD_{ij} - \left(g_iS_iP_j + g_jS_jP_i\right)\label{Eq:dD},
\end{align}
where $\Gamma_T=4\Gamma_z+\Gamma_\uparrow+\Gamma_\downarrow$ is the total molecular linewidth and $S_0=\frac{\Gamma_\uparrow - \Gamma_\downarrow}{\Gamma_\uparrow + \Gamma_\downarrow}$ is the inversion set by the pump.   Note that in writing these equations, we have allowed molecule-dependent coupling strengths $g_i$.

\subsubsection{Homogeneous limit}

We will start by considering the case where $g_i=g$, i.e. where all the molecules couple identically to the light mode.
If we use Eqs. (\ref{Eq:dn}-\ref{Eq:dD}) to adiabatically eliminate everything but the photon occupation $n$ we end up with the quadratic equation for the steady state:

\begin{align}
    0 = & 2\kappa\left[\Gamma_T +\kappa\left(1-\frac{1}{N}\right)\right]\left(\frac{n}{N}\right)^2\\
    & -\left\{\left[\Gamma_T+\kappa\left(1-\frac{1}{N}\right)\right]\left(\Gamma_\uparrow-\Gamma_\downarrow\right)\right.\nonumber\\
      &  \left.-\frac{\kappa\Gamma_T}{N}-\frac{\kappa\Gamma_T\left(\Gamma_T+\kappa\right)}{4g^2N}\left(\Gamma_\uparrow+\Gamma_\downarrow\right)\right\}\frac{n}{N}\nonumber\\
      & -\frac{\Gamma_T\Gamma_\uparrow}{N}.\nonumber
\end{align}

If we make the approximation that $\Gamma_T +\kappa\left(1-\frac{1}{N}\right)\approx\Gamma_T+\kappa$, we can write the solution on the form
\begin{align}
        n = \frac{n_0}{2}\left[\left(\frac{\Gamma_\uparrow}{\Gamma_c}-1\right) +\sqrt{\left(\frac{\Gamma_\uparrow}{\Gamma_c}-1\right)^2 + 4\beta \frac{\Gamma_\uparrow}{\Gamma_c}}\right]\tag{\ref{Eq:nbeta}}.
\end{align}
Introducing the cooperativity $C={4g^2}/{\kappa\Gamma_T}$ we can write the quantities appearing in Eq.~\eqref{Eq:nbeta} as:
\begin{align}
    \Gamma_c &=\frac{NC+1}{NC-1}\Gamma_\downarrow + \frac{NC}{NC-1}\frac{\kappa\Gamma_T}{N(\kappa+\Gamma_T)} ;\\
    n_0 &= \frac{N \Gamma_c}{2\kappa}\frac{NC-1}{NC};\\
    \beta &=\frac{2\kappa\Gamma_T}{N(\kappa+\Gamma_T)}\frac{(NC)^2}{(NC-1)^2\Gamma_c}.
\end{align}

As noted in the main text, the dependence of $\Gamma_T$ on $\Gamma_\uparrow$ means the input-output curve takes a more complex form than discussed by~\cite{Rice_PRA_94}.

\subsubsection{Effects of inhomogeneity}

In reality the electric field strength at the location of the molecules (and thus their coupling to light) will not be identical. This becomes especially important when the molecules are few and the mode volume is very small, as is the case in this paper. We next discuss the effects of such disorder. To do this we reintroduce an $i$-dependence of the coupling strength $g_i$ in the cumulant equations.  In this case, to find the steady state, it is convenient to first solve for $P_i$. This must satisfy the quadratic equation $0=A_iP_i^2+B_iP_i+C_i$ with coefficients that depend on the other $P_j$ via the moments $\Pi^{(n)}=\sum_i g_i^n P_i$.  The coefficients take the form:
\begin{align*}
    A_i&=16g_i^3\\
    B_i&=\Gamma_T(\Gamma_T+\kappa)(\Gamma_\uparrow+\Gamma_\downarrow)\\
      &~+2g_i\bigg[2g_i\Gamma_T\left(1-\frac{4}{\kappa}\Pi^{(1)}\right)
           -(\Gamma_\uparrow-\Gamma_\downarrow)(\tilde g - 2g_i)\\
            &~- 4\left(\Pi^{(2)}+g_i^2\Pi^{(0)}\right)
      \bigg]\\
    C_i&=2g_i\left[\Gamma_T\Gamma_\uparrow-\frac{2\Gamma_T(\Gamma_\uparrow-\Gamma_\downarrow)}{\kappa}\Pi^{(1)}-(\Gamma_\uparrow-\Gamma_\downarrow)g_i\Pi^{(0)}
    \right],
\end{align*}
For a given set $\{g_i\}$ we can solve these equations iteratively. We first guess an initial $\{P_i\}$, then evaluate the moments $\Pi^{(n)}, n=0,1,2$  to find the coefficients $A_i, B_i$ and $C_i$. We then solve the quadratic equation to give a new set $\{P_i\}$. This process can then be iterated to convergence.  We find it necessary to use successive over relaxation to improve the stability of this.

In Fig.~\ref{fig:fig2}(i,j) we show the result of this, where we choose a set of  $g_i$ drawn from a normal distribution of a given mean and variance. In plotting these we in fact adjust the mean of this distribution so as to keep the root mean squared $g$ constant, sincemthe behaviour of strongly-coupled systems typically depends on $\sum_i g_i^2$.  Figure~\ref{fig:fig2}(i) shows a set of input output curves;  one sees that with disorder, each realisation leads to a slightly different curve. Figure~\ref{fig:fig2}(j) is a histogram showing the frequency of slope ratios $\beta$ extracted from these curves, which is therefore the resulting probability distribution of this parameter.


\subsection{Experimental setup}
The details of the setup can be found in Ref. \cite{Ojambati_NatCom_19} and we give a brief description here. For the power dependent measurements, we used a variable neutral density filter to control the intensity of 100~fs pulses at wavelength 640~nm generated from a tunable optical parametric oscillator (SpectraPhysics, Inspire), pumped (SpectraPhysics, Maitai) at 820 $\rm{nm}$ with a repetition rate of 80 $\rm{MHz}$. A brightfield/darkfield microscope objective (Olympus, 100$\times$, numerical aperture = 0.9) focused the attenuated pulses on a single NPoM nanocavity to excite the emitters.
The emitted light was collected in the reflection direction, filtered, and measured using a grating spectrometer (Andor, Shamrock 303i) and an electron multiplying charged-coupled detector (EMCCD, Andor iDus).

The farfield Fourier space imaging was performed by  demagnifying the collected emission by the objective and then focussing it in the focal plane of a lens before the entrance of a monochromator slit. Darkfield spectroscopy was performed on a home-built microscope that illuminates the sample with white light from a halogen lamp (Philips, 100W) at angle of 60$^{\circ}$ through the darkfield microscope. The scattered light was collected through the microscope objective and the scattered spectrum was measured using a fibre-coupled grating spectrometer (OceanOptics). 

\subsection{Sample preparation}
The sample substrate is a Au mirror on a Si wafer that was fabricated using a template-stripping process. Thermal evaporation was used to coat a Si wafer with a 100 nm thick Au film at an average evaporation rate of 0.5 Angstrom/s. Small pieces of another Si wafer (area ~4×4 $\rm{mm}^2$) were glued to the evaporated Au using epoxy (Epo-Tek 377), left overnight at 150$\,^{\circ}$C and then slowly cooled down to room temperature. On applying a slight force, the silicon pieces easily peeled off and were covered by an atomically-flat Au film. A 1 mM  solution of methylene blue (Sigma Aldrich) and a 1 mM solution of cucurbit[7]uril (CB) (Sigma Aldrich) were mixed together, thus allowing the encapsulation of the methylene blue guest molecules inside the CB cavities. A freshly stripped substrate with Au film was submerged in the solution overnight, thoroughly rinsed with de-ionized water, and then blown dry with nitrogen, leaving behind a self-assembled monolayer (SAM) of CB encapsulating a single molecule of methylene blue\, \cite{Chikkaraddy_Nature_16}. Au nanoparticles (BBI Solutions, diameter 60 nm) were drop cast onto the SAM on the Au substrate for 10\,s. The excess Au nanoparticle solution was rinsed with water and blown dry using nitrogen. This leaves sparsely spaced Au nanoparticles deposited on the CB layer
on the Au film, forming the NPoM nanocavity. To control the number of emitters in the nanogap, we diluted the 1 mM methylene blue solution by factors of 2 to 20 and kept the CB concentration fixed at 1mM. Using the surface packing density of CB of 0.24 molecules.nm$^2$~\cite{Horton_PNAS_2020} and a facet size of 15 nm, we estimated the average number of molecules in the nanogap to vary between 2 and 50 for different dilution ratios.

\subsection{Fitting of theoretical curves to experimental data}

In Fig.~\ref{fig:fig2}(d-g) we present results of fitting the theoretical expression Eq.~\eqref{Eq:nbeta} to the experimental input-output curves.   Here we describe the approach used to this fitting.

To perform such a fit, one must relate the theoretical input $\Gamma_\uparrow$ and output $n$ to the experimental power in and power out.  This requires a description of how light couples into and out of the coupling.
The in-coupling is taken to be proportional to the coupling strength squared, $\Gamma_\uparrow \propto g^2 \times$Power density. This is because the external pump couples to the emitters through the same dipole moment and mode profile as controls the light-matter coupling.  As we assume this incoupling is an incoherent absorption process, Fermi's golden rule implies a rate proportional to $g^2$. Including this effect was crucial in producing an accurate fit;  without this factor our fitting led to spurious correlations between fitting parameters.
Out-coupling is simpler, as this involves how the light in the cavity couples to the far-field.  This was assumed to be independent of other parameters and considered as an intensity scaling factor.

With the above assumptions, we then performed a standard least squares fit to find the remaining model parameters.  Some of these parameters are taken as global parameters---common to all experiments; these were $\Gamma_z, \Gamma_\downarrow, \kappa$ and were first optimized (see SI).  The remaining fitting parameters for each NPoM are $g$, an intensity scaling factor, and $N$ which only varies by 20$\%$, due to variations in the number of emitters. The results of the fittings show that there is no correlation between the fitting parameters and that the out-coupling efficiency is linearly proportional to the output intensity, as expected (for more details, see SI).

\section*{Data Availability}
All relevant data in this publication can be accessed at: (to be inserted)

\section*{Acknowledgement} We acknowledge support from EPSRC grants EP/G060649/1, EP/L027151/1, EP/G037221/1,EP/T014032/1, EPSRC NanoDTC, and from the European Research Council (ERC) under Horizon 2020 research and innovation programme PICOFORCE (Grant Agreement No. 883703), THOR (Grant Agreement No. 829067) and POSEIDON (Grant Agreement No. 861950). O.S.O acknowledges the support of a Rubicon fellowship from the Netherlands Organisation for Scientific Research. We thank Rohit Chikkarddy and Rakesh Arul for valuable discussions and Jack Griffiths for making the cartoon in Fig. 1.
\newpage
\onecolumngrid 

\section*{Supplementary Information} 
\renewcommand{\thesection}{S\arabic{section}}
\renewcommand{\thefigure}{S\arabic{figure}}
\renewcommand{\theequation}{S\arabic{equation}}
\setcounter{figure}{0}    
\setcounter{subsection}{0}    
\setcounter{equation}{0}    

\subsection{Sample characterization using darkfield scattering}
\begin{figure*}[h]
    \centering
    \includegraphics[scale = 0.58]{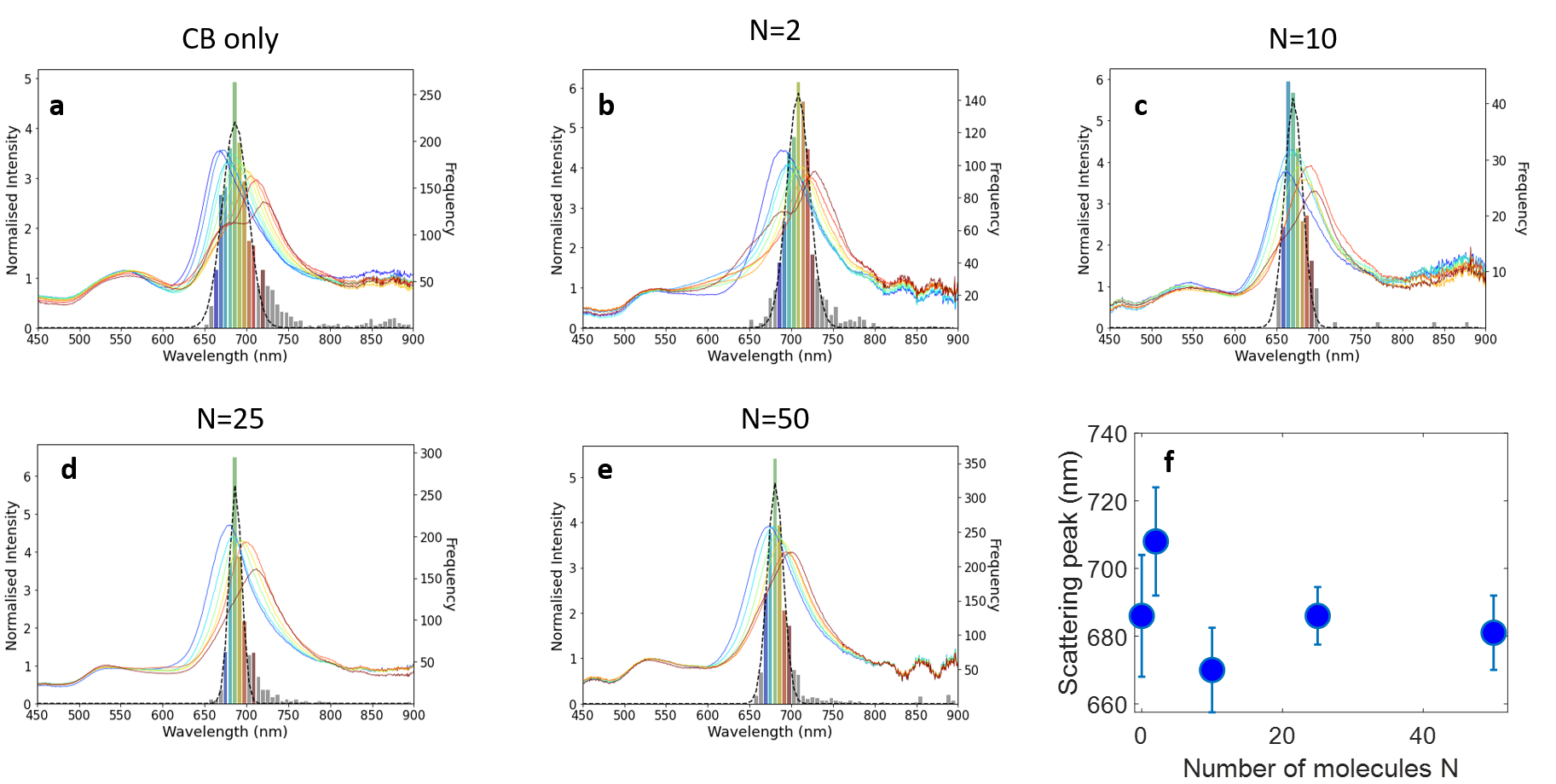}
    \caption{\label{fig:fig_dkfd} (a-e) Scattering spectrum of nanoparticle-on-mirror structures.  Panel (a) shows the case with only cucurbit[7]uril (CB), providing the gap, while panels (b-e) show varying numbers of emitter molecules. (f) Scattering peak \textit{vs} number of molecules $N$.  }
    
\end{figure*}

We prepared several nanoparticle-on-mirror (NPoM) samples with different numbers of methylene blue emitters, that are each encapsulated inside cucurbit[7]uril, defining the nanogap to be 0.9 nm (for sample preparation, see Methods). We characterised the samples using darkfield scattering, which illuminates the NPoM nanostructure at high angles $>60^\circ$ and collected only low-angle scattering from the nanostructure. The high angle illumination helps to reduce specular reflections from the Au film, thus the collected intensity is largely from the nanoparticle. We measured many NPoM samples for different number of emitters ($N$) and plot the histogram of the scattering spectra and representative spectra from each histogram bin (Fig.~\ref{fig:fig_dkfd}, a-e). The scattering spectra show a dominant plasmonic mode that is nearly constant over different $N$ (Fig.~\ref{fig:fig_dkfd} f) with a mean value at 686 nm. Using a plasmonic circuit model, this scattering peak corresponds to a ~ 1~nm nanogap, with a 60~nm Au nanoparticle and assuming a refractive index of 1.6~\cite{Benz_OptExp_15}. This scattering peak is consistent with previous measurements from similar samples~\cite{Chikkaraddy_Nature_16,Horton_PNAS_2020}. 

\subsection{Power dependent measurements}
\begin{figure*}[h]
    \centering
    \includegraphics[scale = 0.5]{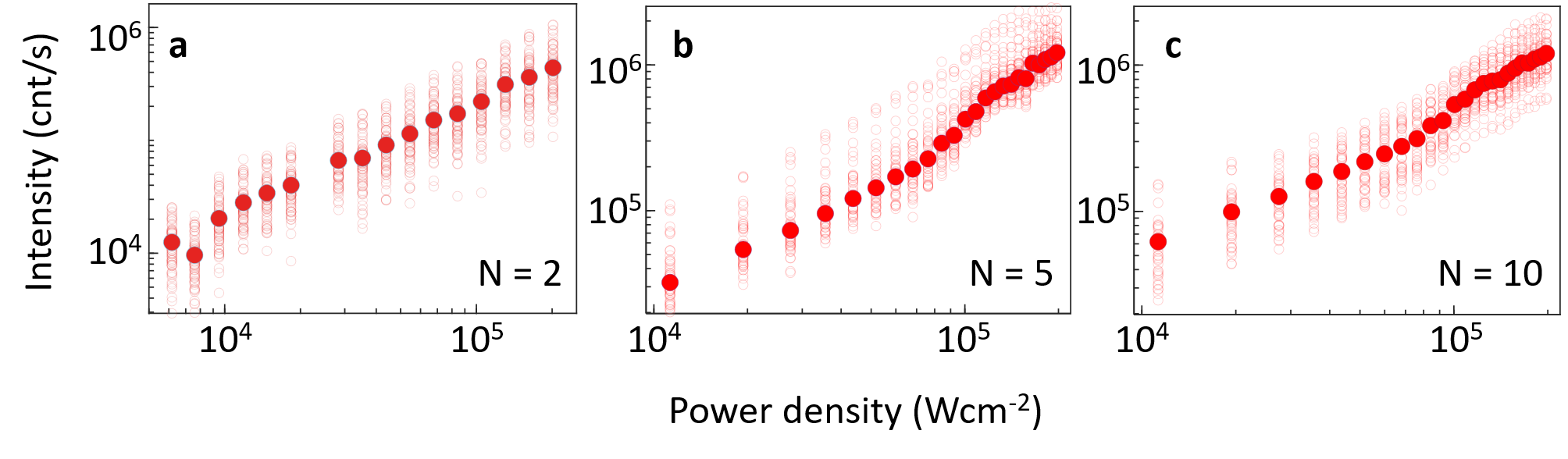}
    \caption{\label{fig:fig_pow_dep} (a-c) Input-output curves of several NPoM samples ($>$ 50, open red circles) each for different number of molecules $N$. The filled red circles are the averages over all the NPoM samples }
    
\end{figure*}

The emission spectra were measured from many NPoMs samples. The measurements were controlled and automated using a custom Python software that used the darkfield scattering image to locate a single NPoM, measure the darkfield spectrum, rotate a variable neutral density filter to control the laser power, and measure the emission spectrum (for optical setup, see Methods). This procedure was repeated for $>$ 50 NPoMs and some examples of the spectrally-integrated intensity are shown in Fig.~\ref{fig:fig_pow_dep}. 

\begin{figure*}[h]
    \centering
    \includegraphics[scale = 0.5]{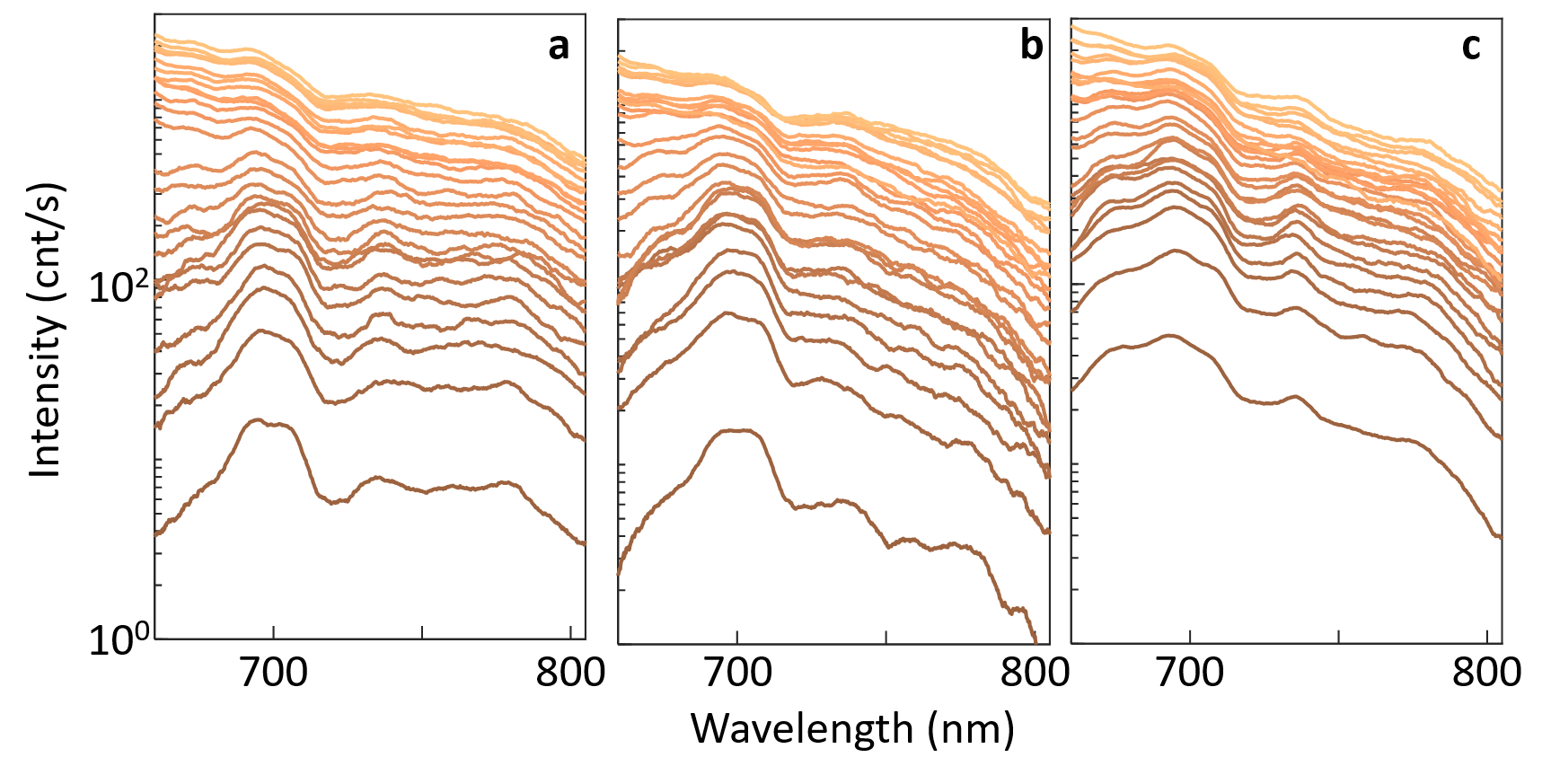}
    \caption{\label{fig:fig_spectra} (a-c) Examples of emission spectra for increasing pump power from three different NPoMs with $N=5$.}
\end{figure*}

\begin{figure}[h]
    \centering
    \includegraphics[scale = 0.5]{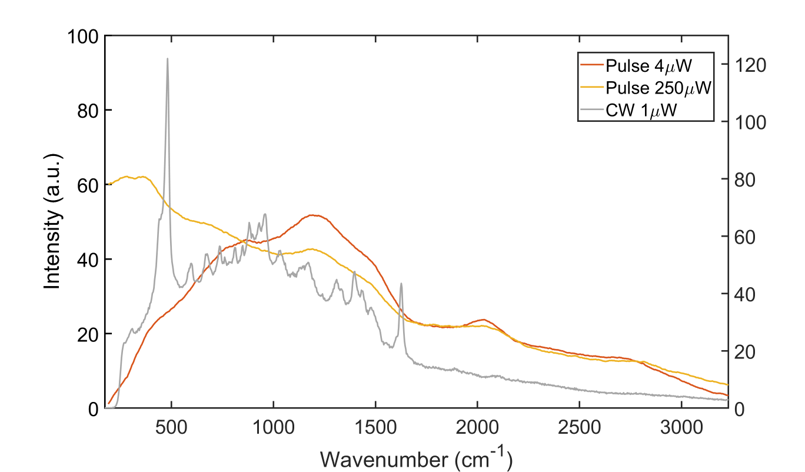}
    \caption{\label{fig:fig_sers} Comparison of surface-enhance Raman scattering using a continuous wave (CW) excitation (wavelength = 633 nm) and emission from NPoM at different pulsed laser powers.}
\end{figure}

As the laser power increases from 4~$\mu$m to 250~$\mu$m, the emission spectra gradually change shape, with the bluer region of the spectra increasing and dominating at the highest powers (Fig.~\ref{fig:fig_spectra}). This effect consistently occurs for several NPoMs with different number of emitters. The increase in the bluer emission implies that emission occurs directly from excited vibronic states instead of following Kasha's rule and emitting from the lowest energy vibronic level. This bluer emission thus suggests a violation of Kasha's rule, resulting from the high Purcell factors in the nanogap, up to $10^5$~\cite{Chikkaraddy_Nature_16,ojambati_efficient_2020}. The high Purcell factor speeds up the radiative decay rate, such that it is faster than the lifetime of the excited vibronic states. A comparison of the emission and surface-enhanced Raman scattering from the NPoM does not show any corresponding peaks between the SERS and emission (Fig~\ref{fig:fig_sers}), implying that the ground state vibrational levels are not involved in the process. It would therefore be interesting in future to investigate the impact of the various vibronic states on the light-matter interactions in this lasing regime~\cite{zhao_impact_2020}.

\subsection{Fitting results}

\begin{figure*}
    \centering
    \includegraphics[scale = 0.55]{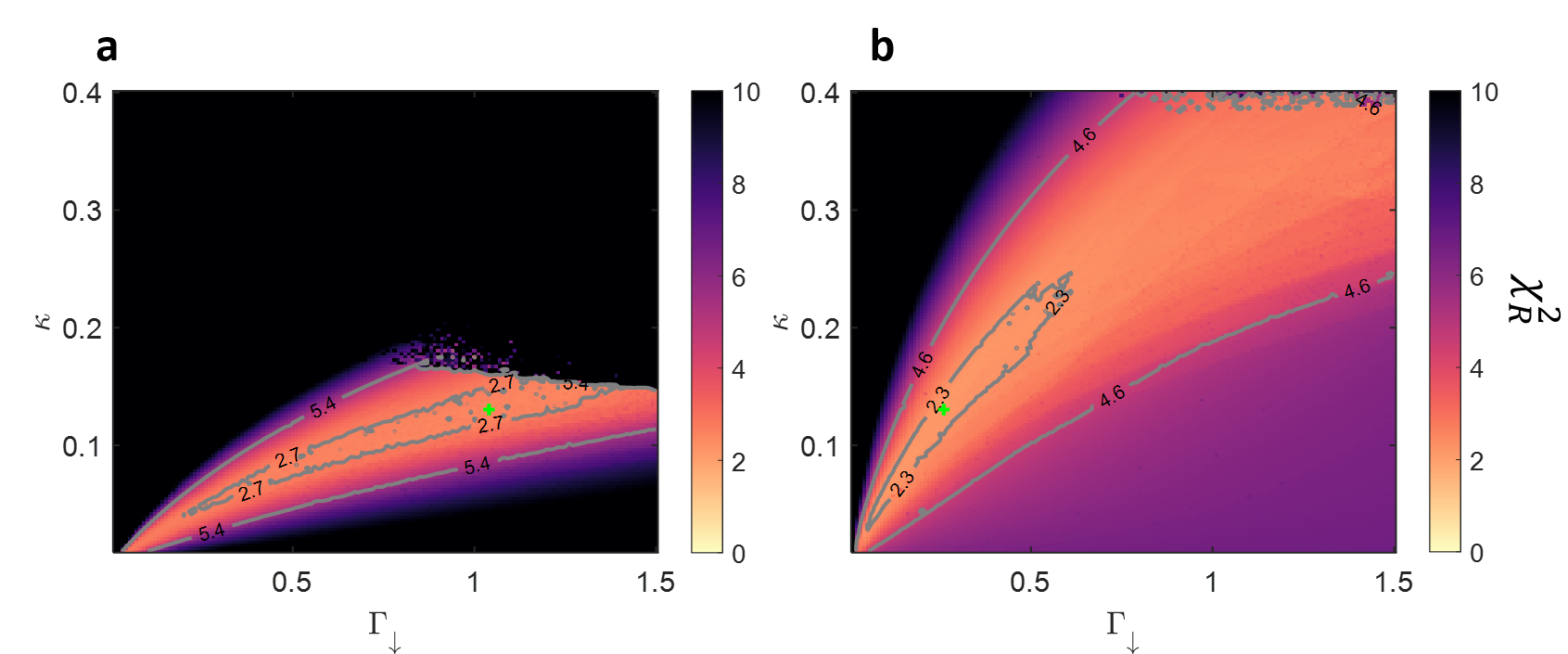}
    \caption{\label{fig:fig_maps} Reduced $\chi^2$ maps obtained by optimizing the values of $\Gamma_\downarrow, \kappa$ for a fixed value of $\Gamma_z=10$ for N=5 (a) and N=10 (b).The green cross shows the minimum value of the $\chi_R^2$ and the grey curves are the contour lines of the 1$\sigma$ and 2$\sigma$ confidence intervals.  As discussed in the text, the units of these parameters are arbitrary, and only (dimensionless) ratios matter for the fitting.    }
\end{figure*}

As described in the Methods section, we fit the theoretical result to the experimental data to determine relevant parameters. 
We divide parameters into three classes:
\begin{enumerate}
    \item[A.] Parameters which we fix globally over all NPOMs.  In this class is $\Gamma_z$, which is assumed an intrinsic property of the molecules.
    \item[B.] Parameters which are constant over NPOMs but allowed to vary according to the concentration of emitters (and thus target $\bar{N}$).  In this class are $\Gamma_\downarrow, \kappa$.
    \item[C.] Parameters which vary between each NPOM.  In this class is the value $g$ and an output intensity scaling which we denote $S_{\text{out}}$.  This output scaling relates the measured output counts $C_{\text{out}}$ to the theoretical photon number $n$ vias $C_{\text{out}} = S_{\text{out}} n$.  In addition, we allow the value $N$ to vary, within 20\% of the target $\bar{N}$, to account for variation in the number of emitters actually coupled.
\end{enumerate}
To perform the fitting, for each value of parameters in class B we optimise over the parameters in the class C (by simple least squares), and then calculate a reduced $\chi_R^2$ value from the match of the theory to the data when using those class-B parameters.  We repeat this over a range of values of class-B parameters.  This gives us a map of $\chi_R^2$
(see Fig.~\ref{fig:fig_maps} for examples) from which we choose the optimal class-B parameters.  
We then use these parameters, along with the standard least squares fit of parameters in class C.
Plotting the $\chi_R^2$ map allows us to identify how tightly confined are the class-B parameters.
Following a Gaussian error model, 68~$\%$ of fits should have values of $\chi_R^2$ satisfying 
\begin{equation}
\chi_R^2 \leq \tilde{\chi}_R^2 =  \chi_{R, min}^2 + \frac{1}{k_{\rm{eff}}}\Delta\,
\end{equation}
and finding the area in parameter space defined by this inequality allows us to estimate errors.
Here, $k_{\rm{eff}}$ and $\Delta$ are the reduced number of values and $p$-value from a $\chi^2$ distribution table. This area is shown by grey contour in Fig.~\ref{fig:fig_maps}.

In performing this fitting, we take the theoretical parameter $\Gamma_\uparrow = g^2 \times (Power density)$, where Power density is the measured value in units of $\text{W} \text{cm}^{-2}$.  The theoretical behaviour depends only on dimensionless ratios of the parameters $\Gamma_\uparrow, \Gamma_\downarrow, \Gamma_z, \kappa, g$, in order to produce the dimensionless photon number $n$.  As such, the somewhat unusual units for $\Gamma_\uparrow$ do not cause a problem.  Physical units could be restored by introducing an extra scaling parameter between power density and $\Gamma_\uparrow$, however since the theoretical fits depend only on ratios of parameters, the fit would be degenerate in this parameter.

A subset of the measured input-output curves appear to show different behaviour, with a saturation of the output power occurring below the lasing threshold.  As such, these datasets cannot be fitted to the theoretical model.  The origin of this different form of behaviour may be physical or instrumental, which needs further investigation. The fit to these datasets are considered invalid, since the extracted fitting parameters give erroneously high values. Thus, we consider only fits where $\chi_R^2 \leq \tilde{\chi}_R^2$. These fits account for $<$ 30\% of the entire dataset shown in Fig. S2.

\begin{figure*}
    \centering
    \includegraphics[scale = 0.55]{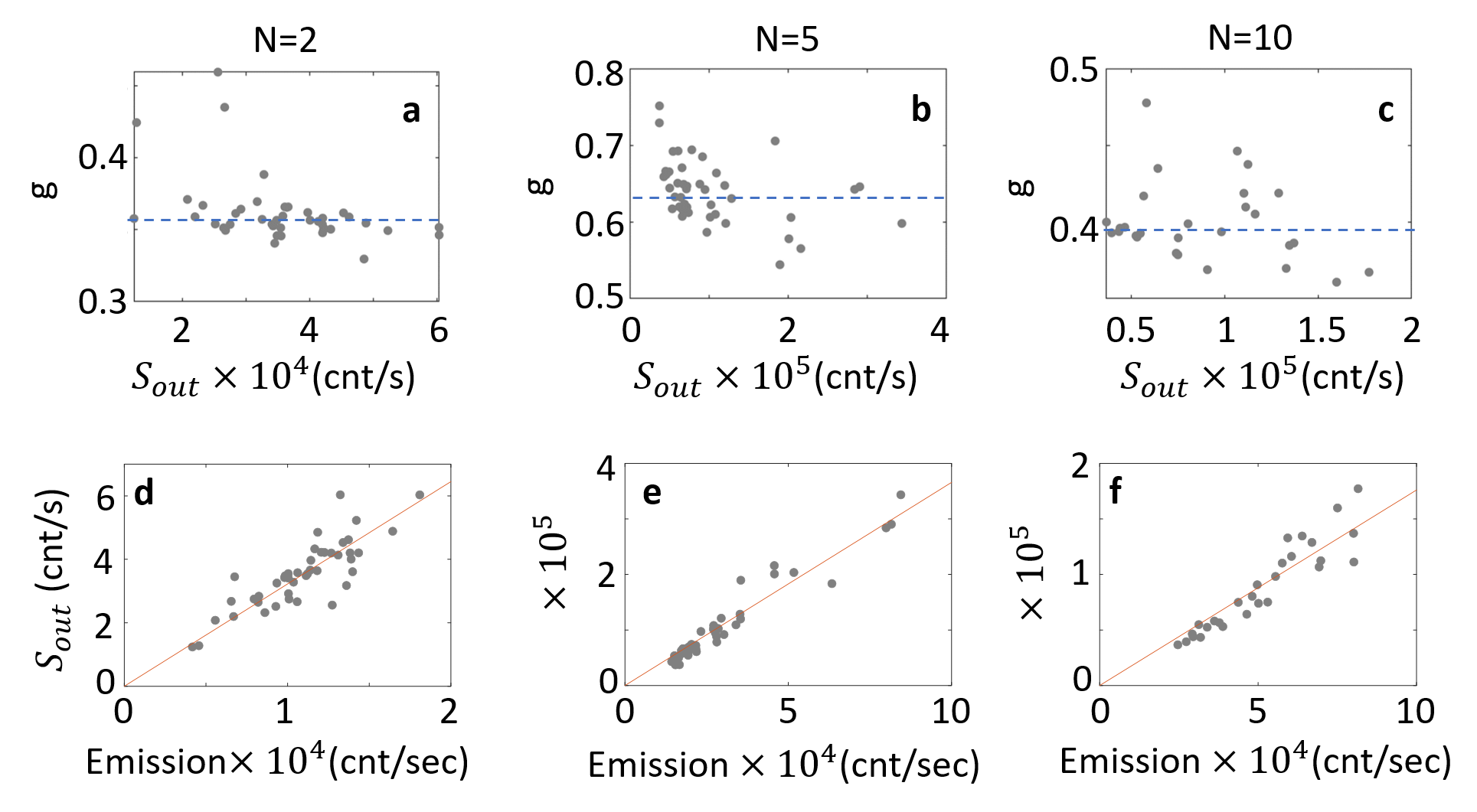}
    \caption{\label{fig:fig_g_vs_I} (a-c) The extracted coupling strength $g$ $vs$ intensity scaling, showing uncorrelation between the parameters as indicated by the guide-to-the-eye blue line. (d-f) Intensity scaling $vs$ emission below threshold and the orange curve is a linear fit.}
\end{figure*}

There is no correlation between the fitting parameters in the class C [Figs.~\ref{fig:fig_g_vs_I}(a-c)], for $\chi_R^2 \leq \tilde{\chi}_R^2$. This lack of correlation is expected, since the coupling between emitters and cavity mode, and the out-coupling of the cavity mode to the detector should be independent parameters. 
There is however a  correlation between the outcoupling intensity scaling and the emission below threshold [Figs.~\ref{fig:fig_g_vs_I}(d-f)], which implies that an efficient outcoupling efficiency of NPoM cavity results into a high emission intensity.

\subsection{Shape of input-output curve}

As mentioned in the main text, the shape of the input-output curve is not fixed when one goes beyond the weak-coupling laser theory of Ref.~\cite{Rice_PRA_94}.  In the weak coupling limit, the parameter $\beta$ fully determines the shape of the curve, while other parameters merely rescale the photon number or input power.  In contrast, for strong coupling---particularly combined with inhomogeneous coupling strengths $g_i$---the shape can vary.  

 \begin{figure*}
     \centering
     \includegraphics[scale = 0.6]{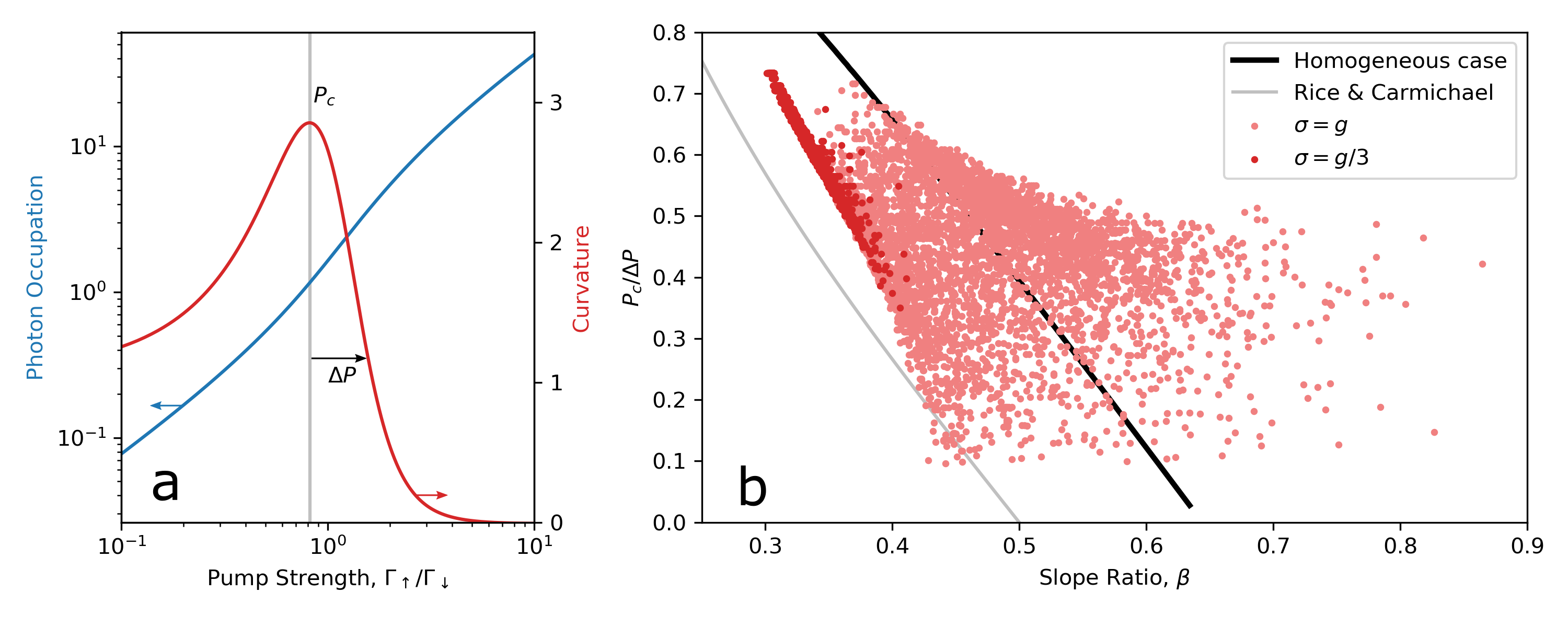}
   \caption{\label{fig:fig_shape}Variation of shape of input-output curve.  (a) Definition of transition width, extracted from the ratio of critical power $P_c$ as defined by peak curvature, to right half-width half-maximum of curvature $\Delta P$. Here curvature is the second derivative of input--output curve.  (b) Relation between transition width, $P_c/(\Delta P)$, and slope-ratio $\beta$.  The thick black line indicates the homogeneous limit of our model (no disorder in couplings.  The gray line instead shows the weak-coupling formula of Ref.~\cite{Rice_PRA_94}.  The points correspond to calculations with disordered couplings $g_i$, choosing $g_i$ drawn from a distribution with standard deviations as indicated in the legend.  Notably, with this disorder, there is no longer a one-to-one relation between transition width and slope ratio:  i.e. the shape of the input--output curve varies. Parameters used (matching Fig.~\ref{fig:fig2}(i,j)): $\Gamma_z=\Gamma_\downarrow, \kappa=1.74\Gamma_\downarrow$. In the homogeneous case, a range of values of ${g}$ are considered to map out the range of $\beta$. In the inhomogeneous cases, we fix the root mean square of $g_i$ to be $1.5\Gamma_\downarrow$, and sample distributions with widths as indicated in the legend. }
 \end{figure*}

One way to characterise the variation of input-output curve shape is to consider the relationship between slope ratio $\beta$ and the `width` of the transition.  As discussed in the main text, we define $\beta$ by the ratio of slopes below and above threshold. We define transition width by considering the curvature of the input-output curve;  the curvature should vanish in both the normal and lasing phases (where output is proportional to input).  As such, we can use the peak curvature to numerically locate the critical power $P_c$, and the half-width half-maximum of curvature as a transition width $\Delta P$.   We use the right half-width to avoid problems when the transition width becomes comparable to the critical power.    With these defined, we can calculate a dimensionless parameter $P_c/(\Delta P)$ which is a characteristic of the lasing transition.  Figure~\ref{fig:fig_shape} shows the relation between the dimensionless parameters $\beta$ and  $P_c/(\Delta P)$.   

If the input-output curve were of fixed shape, there would exist a one-to-one relationship between $\beta$ and  $P_c/(\Delta P)$.  We see first that the relationship predicted by our model without disorder differs from that of Ref.~\cite{Rice_PRA_94}.  We note further (not shown) that this relationship is now parameter dependent, and would change depending on, e.g. $\Gamma_z/\Gamma_\downarrow, \kappa/\Gamma_\downarrow$.  More importantly,  the one-to-one relation clearly breaks down when we consider inhomogeneous couplings, with varying $g_i$.  This demonstrates that the input-output curve does not in general have a fixed shape.

\end{document}